\documentstyle[12pt]{article}

\author{Piret Kuusk\\
Institute of Physics, University of Tartu,\\ 
Riia 142,  Tartu 51014, Estonia, \\
e-mail: piret@fi.tartu.ee}

\title{GEODESIC MULTIPLICATION AND QUANTUM KINEMATICS IN A 
NEWTONIAN SPACETIME
}
\date{\mbox{}}

\begin{document}
\maketitle
\begin{abstract}
We consider a quantum test particle in the background of a Newtonian
gravitational field  in the framework of  Cartan's
 formulation of nonrelativistic spacetime.
We have proposed a novel quantization of a point particle which amounts to
introducing its position operators as multiplication with the corresponding
Riemann normal coordinates and momentum operators as infinitesimal
right translation operators determined by geodesic multiplication of
points of the spacetime. We present detailed calculations for the 
simplest model of a two-dimensional Newtonian spacetime.
\end{abstract}

%%%%%%%%%%%%%%%%%%%%%%%%%%%%%%%%%%%%%%%%%%%%%%%%%%%%%%%%%%%%%%%%%%%%%%%%%%%
{\bf 1. Introduction.}
Let us consider a 4-dimensional differentiable  manifold $M$ with
a symmetric affine connection  (a spacetime with vanishing torsion).
Every reasonable neighbourhood $M_e \in M$ can be equipped with a 
nonassociative binary operation called the geodesic multiplication
\cite{Kikkawa, Akivis, Sabinin, kopjmp}.

Its infinitesimal left and right translations can be used to define the
(geodesic) momentum operators  of a quantum test particle in the curved 
spacetime  \cite{kopcqg, plb}. 
The corresponding commutation relations can be taken as the quantum 
kinematic algebra. 
It coincides with the usual canonical Poisson algebra (Weyl's kinematics) 
only in the case of the flat spacetime.
In the present paper, detailed calculations are performed for the 
spacetime which describes a nonrelativistic Newtonian gravitational field. 

%%%%%%%%%%%%%%%%%%%%%%%%%%%%%%%%%%%%%%%%%%%%%%%%%%%%%%%%%%%%%%%%%%%%%%%%%%%
{\bf 2. Geodesic multiplication and translations.}
The local geodesic multiplication of points $x,y \in M_e$ is defined 
\cite{Kikkawa, Akivis} by
\begin{equation}
xy \equiv L_x y \equiv R_y x = \bigl(\exp_y\circ\tau^e_y
\circ\exp^{-1}_e\bigr) x. \label{geodkorr}
\end{equation}
Here we have used the exponential mapping 
$T_e M \rightarrow M_e : X \mapsto x= \exp_e X =x(1;X)$ and
the parallel transport mapping $\tau^e_y : T_e M \rightarrow T_y M$
along a geodesic $y(s)$ emerging from $e \in M_e$.
The local geodesic multiplication can be constructed in every 
neighbourhood $M_e$ where all required exponential mappings and
parallel transport operations are well defined.
$M_e$ is a finite region of $M$, where geodesics emerging 
from $e$ do not intersect and which does not contain singular points.

By introducing the local geodesic multiplication, $M_e$ can be seen to be a 
binary algebraic system called the local geodesic loop 
\cite{Kikkawa,  Akivis, Sabinin}. 
The point $e\in M$ is the unit element of the loop $M_e$ and
the (left) inverse $x^{-1}_L$ of $x\in M_e$ is defined by $x^{-1}_Lx=e$.

In general, the geodesic multiplication need not be commutative and 
associative \cite{Akivis}. 

The local geodesic multiplication (\ref{geodkorr}) determines the following
left ($L$) and right ($R$) infinitesimal translation matrices:
\begin{eqnarray}
(xy)^\mu &=& y^\mu+L^\mu_\nu(y)x^\nu+\ldots\,,\qquad\, 
             L^\mu_\nu(y)\equiv{\partial(xy)^\mu\over
                 \partial x^\nu}\Big|_{x=e},          \label{left} \\
(xy)^\mu &=& x^\mu+R^\mu_\nu(x)y^\nu+\ldots\,,\qquad
             R^\mu_\nu(x)\equiv{\partial(xy)^\mu\over
                 \partial y^\nu}\Big|_{y=e}.          \label{right}
\end{eqnarray}
Note that the lower indices of matrices $L^\mu_\nu$, $R^\mu_\nu$
in fact belong to the tangent space $T_eM$. In the
following let us denote tangent space (flat) indices with letters 
from the beginning of the alphabets. 
Now two local frame fields can be introduced
\begin{equation}
L_\alpha(x)\equiv L^\mu_\alpha(x){\partial_\mu},\qquad
R_\alpha(x)\equiv R^\mu_\alpha(x){\partial_\mu}.       \label{reeperid}
\end{equation}
It is well known that for two vector fields their commutator is again a
vector field. We know that $L_\alpha(x)$ and $R_\alpha(x)$ 
are frame fields, so
it is  quite natural to define the structure functions
$\lambda_{\alpha \beta}^\gamma(x)$ and $\rho_{\alpha \beta}^\gamma(x)$ by
\begin{equation}
[L_\alpha(x),L_\beta(x)]=-\lambda_{\alpha \beta}^\gamma(x)L_\gamma(x),
\label{AL}
\end{equation}
\begin{equation}
[R_\alpha(x),R_\beta(x)]=+\rho_{\alpha \beta}^\gamma(x)R_\gamma(x).
\label{AR}
\end{equation}
In general, the structure functions  
$\lambda_{\alpha \beta}^\gamma(x)$,  $\rho_{\alpha \beta}^\gamma(x)$
do not coincide. 

%%%%%%%%%%%%%%%%%%%%%%%%%%%%%%%%%%%%%%%%%%%%%%%%%%%%%%%%%%%%%%%%%%%%%%%%%%%%%
{\bf 3. Geodesic translations in the Riemann normal coordinates.}
Let us introduce  the Riemann normal 
coordinates $y^\alpha$ for which  $\Gamma^\alpha_{\beta\delta}(e)=0$.
In these coordinates,  
expansions of $R^\alpha_\beta(y)$ and $L^\alpha_\beta(y)$  
  can be found  
using Eqs.~(\ref{left}), (\ref{right}) and the geodesic multiplication 
formula of Akivis \cite{Akivis}
\begin{equation}
(zy)^\alpha = z^\alpha + y^\alpha - {1 \over 2}
\Gamma^\alpha_{\beta \gamma , \delta} (e) z^\beta z^\gamma y^\delta - 
{1\over 2}
\Gamma^\alpha_{\beta (\gamma , \delta)} (e) z^\beta y^\gamma y^\delta
+ \ldots \,.    \label{akivis}
\end{equation}
We get
\begin{eqnarray}
R^\alpha_\beta(y) &=& \delta^\alpha_\beta - {1 \over 2}
\Gamma^\alpha_{\beta \gamma , \delta} (e) y^\beta y^\gamma + \ldots\, 
\label{R}\\ 
L^\alpha_\beta(y) &=& \delta^\alpha_\beta - {1\over 2}
\Gamma^\alpha_{\beta (\gamma , \delta)} (e)  y^\gamma y^\delta +
\ldots  \label{L}
\end{eqnarray} 
Now we can introduce local frame fields $L_\alpha(y)$ and $R_\alpha(y)$. 
Their structure functions $\lambda_{\alpha \beta}^\gamma(y)$ 
and $\rho_{\alpha \beta}^\gamma(y)$ have expansions 
\cite{kopcqg, paalpre}
\begin{eqnarray}
\lambda_{\alpha \beta}^\gamma(y)&=&-{R^\gamma}_{[\alpha \beta]\delta}(e) 
y^{\delta} +\ldots\, ,  \label{lambda}\\
\rho_{\alpha \beta}^\gamma(y)   &=&{R^\gamma}_{\delta[\alpha \beta]}(e)
y^{\delta} +\ldots\, .  \label{rho}
\end{eqnarray}
The commutator of two frame fields can also be calculated 
\cite{kopcqg, paalpre}
\begin{equation}
[L_\alpha (x), R_\beta(x)]= {1 \over 2} {R^\kappa}_{\alpha \delta \beta} (e)
y^\delta \partial_\kappa + \ldots .
\end{equation}
Here ${R^\gamma}_{\alpha \beta \delta}(e)$ denote components of the
curvature tensor  at the unit element $e $. We use  definitions and 
sign conventions  given in \cite{mtw}. 

%%%%%%%%%%%%%%%%%%%%%%%%%%%%%%%%%%%%%%%%%%%%%%%%%%%%%%%%%%%%%%%%%%%%%%%%%%%%%
{\bf 4. Structure of a nonrelativistic spacetime.}
Let us consider a four--dimensional differentiable manifold $V$ with
an affine connection and a vanishing torsion.
Following \cite{cartan}, let us introduce the absolute
time by imposing two conditions.

1. There exists a function $t(x^\mu)$ (the absolute time) and therefore
1--form $dt$ that is exact, 
$$
dt \equiv t_\mu dx^\mu = t_{,\mu} dx^\mu, \quad d \wedge dt = 0 . 
$$

2. 1--form $dt$ is covariantly constant, 
$$
\nabla_u dt =
\nabla_u (t_\mu dx^\mu) \equiv u^\nu (t_{\mu, \nu} - \Gamma^\sigma_{\mu \nu}
t_\sigma) dx^\mu = 0 
$$
for all vectors $u \in T V$.

For introducing the absolute space we have  proposed \cite{hk, ph}  
to impose  the 
condition of vanishing curvature of three--hyperplanes $t = const$
with coordinates $x^i$ $(i= 1,2,3)$,
${R^a}_{bcd}(x^i) = 0. $
It is slightly less restrictive than the corresponding conditions 
${R^\alpha}_{b \mu \nu} = 0$,  ${R^\alpha}_{\beta mn}=0$ 
given in \cite{mtw}. 

Let us introduce basis 1--forms $\omega^\alpha$. 1--form $\omega^0$
can be chosen in the direction of $dt$, $\omega^0 \sim dt$. The
scalar product between basis 1-forms is defined as
\begin{equation}
\omega^\alpha \cdot \omega^\beta = e^{\alpha \beta} \equiv
\left(\matrix{ 0&0&0&0\cr  0&1&0&0\cr 0&0&1&0\cr 0&0&0&1\cr}\right).
\label{smetric}
\end{equation}
Now the relation $\rho \cdot dt = 0 $ holds for every 1-form $\rho$.
Singular metric $e^{\alpha \beta}$ is supposed to be covariantly
constant, $\nabla_\mu e^{\alpha \beta} = 0$. 

Let us introduce  also connection 1--forms 
${\omega^\alpha}_\beta \equiv \Gamma^\alpha_{\beta \delta} \omega^\delta$.
By definition the following relation holds:
$$ 
\nabla_\beta \omega^\alpha = - \Gamma^\alpha_{\delta \beta} 
\omega^\delta. 
$$
Applying the covariant differentiation operator $\nabla_\beta$ 
to  condition (\ref{smetric}) we get
 the following  properties of the connection 1--forms:
$$  
{\omega^a}_d \delta^{bd} + {\omega^b}_d \delta^{ad}= 0,
$$
$$  {\omega^0}_a = 0. 
$$
We don't get any conditions for connection 1--forms 
${\omega^i}_0$ because of the special form of singular metric
$e^{\alpha \beta}$.

We can introduce also a dual basis consisting of vectors 
$e_\alpha = {e_\alpha}^\mu {\partial \over \partial x^\mu}$,
$<\omega^\alpha, e_\beta> = \delta^\alpha_\beta$. 
But because of the singularity of metric $ e^{\alpha \beta} $
one cannot introduce the corresponding 
scalar  product between vectors $e_\alpha$.

%%%%%%%%%%%%%%%%%%%%%%%%%%%%%%%%%%%%%%%%%%%%%%%%%%%%%%%%%%%%%%%%%%%%%%%%
{\bf 5. Nonrelativistic spacetime with a Newtonian gravitational field.} 
In a nonrelativistic spacetime where all the abovementioned conditions hold
there exists a preferred coordinate system (Galilean coordinates $x$)
where the basis 1--forms are simply  differentials of holonomic
coordinates 
$ \omega^\mu = dx^\mu$. In particular,  
$ \omega^0 \equiv dx^0 = dt$.

Let us consider a nonrelativistic spacetime specified by  
a Newtonian gravitational potential $\Phi(x)$. 
Let us identify  trajectories of freely falling
test particles 
\begin{equation}
{d^2 x^j \over dt^2} + {\partial \Phi \over \partial x^j} = 0 \label{Newton}
\end{equation}
with geodesic lines of a curved spacetime
\begin{equation}
{d^2 x^j \over dt^2} + \Gamma^j_{\mu \nu} {dx^\mu \over dt} {dx^\nu \over dt}
=0.
\end{equation}
As a result, the Newtonian spacetime has  a nontrivial affine connection 
and a nonvanishing curvature tensor \cite{mtw, cartan} 
\begin{equation}
 \Gamma^j_{00}=  {\partial \Phi \over \partial x^j} \,, \qquad 
{R^j}_{0k0}=- {R^j}_{00k} = {\partial^2 \Phi \over \partial x^j \partial x^k},
\label{koverus}
\end{equation}
all other components vanish.

Let us introduce the Riemann normal coordinates $\{t_R, y^i\}$
with the origin of coordinates $e^\mu = 0$. In the linear approximation 
in gravitational potential $\Phi$ the transformation formulas read
\begin{eqnarray}
t &=& t_R ,   \\
x^j &=& y^j - {1 \over 2} {\partial \Phi (0) \over \partial x^j} t_R^2
- {1 \over 6} { \partial^2 \Phi (0) \over \partial x^j \partial x^k} 
t_R^2 y^k + \ldots . \label{GR}
\end{eqnarray}
The Galilean absolute time $t$ does not transform and in the following 
we need not specify the
coordinates to which it belongs.

Geodesic lines emerging from $e=0$ -- trajectories of freely falling 
test particles -- are now given by equations
\begin{equation}
y^i (t) = v^i(0) t\, \label{geodline}
\end{equation}
where  $v^i(0)= const$ are initial velocities, $v^i(0)={\dot y}^i(0)
={\dot x}^i(0)$.  Information about the gravitational field is
encoded in the transformation (\ref{GR}) from  Galilean to 
 Riemannian coordinates: according to Eq. (\ref{Newton}), 
 $-  \partial \Phi(x) / \partial x^j$ is the gravitational force
 acting upon a test particle with a unit mass.

%%%%%%%%%%%%%%%%%%%%%%%%%%%%%%%%%%%%%%%%%%%%%%%%%%%%%%%%%%%%%%%%%%%%%%% 
{\bf 6. A  two-dimensional model.}
For simplicity, let us consider a two-dimensional Newtonian spacetime 
with an absolute time and one space coordinate. 
In the Riemann normal coordinates $\{t, y\}$,  nonvanishing
connection coefficients have the following expansions in $\{t,y\}$ and in
$\Phi$:
\begin{equation}
\Gamma^y_{00} = { 2 \over 3} Cy + \ldots, \qquad  \Gamma^y_{y0} =
- { 1 \over 3} C t+ \ldots, \label{conn}
\end{equation}
where we have introduced a notation
$   \partial^2 \Phi (0) / \partial x \partial x \equiv C $. 
From Eqs. (\ref{R}), (\ref{L}) the right and the left translation 
matrices can
be calculated. Expansions of the right and the left local frame fields 
(\ref{reeperid})  read:
\begin{eqnarray}
R_0 = R^s_0 \partial_s &=& \partial_t + \bigl({1 \over 3} 
Cty  +\ldots \bigr) \partial_y , \\
R_y = R^s_y \partial_s &=& (1 - {1 \over 3} Ct^2+\ldots)\partial_y; \\
L_0 = L^s_0 \partial_s &=& \partial_t - \bigl({1 \over 6} Cty
+\ldots\bigr) \partial_y, \\
L_y = L^s_y \partial_s &=& (1 + {1 \over 6} Ct^2+ \ldots)\partial_y.
\end{eqnarray}
The corresponding  structure functions are
\begin{equation}
\rho^y_{0y} = - Ct+ \ldots \, \qquad \lambda^y_{0y} =- {1 \over 2 } Ct+\ldots
\end{equation}
and
\begin{eqnarray}
{[}L_0, R_0 {]} &=& \bigl( {1 \over 2} Cy + \ldots \bigr) \partial_y, \\
{[}L_0, R_y{]}  &=& \bigl( - {1 \over 2} Ct + \ldots \bigr) \partial_y,\\
{[}L_y, R_0{]}  &\approx& 0, \qquad   [L_y, R_y] \approx 0.
\end{eqnarray}
Note that in the lowest approximation,  $\rho$ and $\lambda$ depend 
only on time $t$.

%%%%%%%%%%%%%%%%%%%%%%%%%%%%%%%%%%%%%%%%%%%%%%%%%%%%%%%%%%%%%%%%%%%%%%%%
{\bf 7. A quantum test particle in a curved spacetime.}
Let us consider a quantum test particle  
moving in a curved spacetime \cite{kopcqg, plb}.
Let us introduce an action (on scalar valued functions)
 of the position operators ${\bf y}^\alpha$ 
 as multiplication with the Riemann normal 
coordinates $y^\alpha$. Then we have 
\begin{equation}
{[}{\bf y}^\alpha,{\bf y}^\beta{]}=0. \label{kan1} 
\end{equation}
We have proposed to define (geodesic) momentum operators ${\bf p}_\alpha$ via 
infinitesimal right geodesic translations
 ${\bf p}_\alpha(y)=-i\hbar R^\beta_\alpha(y)\partial_\beta$.
Then,
\begin{equation}
 {[}{\bf y}^\alpha,{\bf p}_\beta(y){]}
                = i\hbar R^\alpha_\beta(y)             \label{kan2}
\end{equation}
and 
\begin{equation}
{[}{\bf p}_\alpha,{\bf p}_\beta{]}  
=-i\hbar\rho_{\alpha \beta}^\delta {\bf p}_\delta.  \label{kan3}
\end{equation}
Commutators (\ref{kan1})--(\ref{kan3}) constitute a
modification of canonical quantum commutation relations.

%%%%%%%%%%%%%%%%%%%%%%%%%%%%%%%%%%%%%%%%%%%%%%%%%%%%%%%%%%%%%%%%%%%%%%%
{\bf 8. Modified canonical formalism for a test particle 
in a two-dimensional Newtonian spacetime.}
To clarify the physical and geometrical meaning of the quantum commutation
relations (\ref{kan1})--(\ref{kan3}), let us consider
the corresponding modified canonical formalism in the classical mechanics
for a test particle moving in a two-dimensional Newtonian spacetime.
Let us replace quantum commutators with classical brackets
as follows:
\begin{equation}
{ {[}{\bf y}^\alpha, {\bf p}_\beta{]} \over i \hbar}
 \rightarrow [y^\alpha, p_\beta], \qquad {{\bf p}_\beta \over i\hbar} 
 \rightarrow p_\beta.
\end{equation}
Using  results of Sec. 6,  nonvanishing brackets in the lowest
approximation are  
\begin{eqnarray}
{[}y, p_y{]} &=& 1 - {1 \over 3} C t^2 , \label{komm1}\\
{[}t, p_0{]} &=& 1, \label{komm2}\\
{[}y, p_0{]} &=& {1 \over 3}  Cty  \label{komm3}
\end{eqnarray}
and
\begin{equation}
[p_y, p_0] = -Ctp_y  . \label{kommp}
\end{equation}

Let us introduce the
canonical momentum vector $ p_k^{can} $ with brackets
\begin{equation}
[y^i, p_k^{can}] = \delta^i_k. \label{pcan}
\end{equation}
Then brackets (\ref{komm1})--(\ref{komm3})  
can be considered as defining the
components of geodesic momentum vector   $ p_k $
in terms of canonical momentum $ p_k^{can} $ 
as follows:
\begin{eqnarray}
p_y &=& p^{can}_y   (1 - {1 \over 3} Ct^2 ) , \\   
p_0 &=& p^{can}_0 +  {1 \over 3} Cty  p^{can}_y .
\end{eqnarray}
Note that the canonical momentum of a test particle
is the main part of its geodesic momentum.

Let us try to find a physical interpretation
of the geodesic momentum. In our two-dimensional model, 
the spacetime can be considered as consisting of
a family of geodesic lines. 
Let us denote $ v(0)= v $, then $v$ acts as a    
a parameter which enumerates the geodesics.   
The time $t$ is a canonical parameter along a geodesic.
Now  $y^i = \{t,\quad vt\}$, $t>0$ can be considered as a change of
coordinates where the new coordinate lines are $g^i(t)= y^i(t,\quad v=const)$
and $h^i(v) = y^i (t=const, \quad v)$. The corresponding tangent
vectors are
\begin{equation}
U^i = {dg^i(t) \over dt} = \{1, \quad v\}, \qquad 
V^i = {dh^i(v) \over dv} = \{0, \quad t\}.
\end{equation}
Using the expressions for connection coefficients (\ref{conn}),
let us calculate the following absolute derivatives ${D \over Dt}$ 
along lines $y^i =g^i(t)$ -- the trajectories of freely falling particles 
(\ref{geodline}):
\begin{eqnarray}
{DU^i \over Dt} &\equiv& {dU^i \over dt} + \Gamma^i_{kl}U^k {dy^l \over dt}
 = 0 , \label{DU}\\
{DV^i \over Dt}&=& \{0, \quad 1-{1 \over 3} Ct^2 \}, \label{VV}\\
{D^2 V^i \over Dt^2} &=& \{0, \quad -Ct\}. \label{D2V}
\end{eqnarray}
Eq. (\ref{DU}) confirms that lines $y^i =g^i(t)$ are geodesic lines.
Vector $V^i dv$ is a vector joining a geodesic to a neighbouring
geodesic.  The rate of change of $V^i$ along the geodesic $g(t)$ is
given by  the geodesic deviation equation which now reads 
\begin{equation}
{D^2 V^1 \over Dt^2} = -Ct.
\end{equation} 
In the case of a test particle with a unit mass, canonical commutation
relations (\ref{pcan}) allow us to identify $p_y^{can} = {\dot y} = v$
and write $y = vt = p_y^{can} V^1$. As a result we get in the lowest
approximation
\begin{eqnarray}
{Dy \over Dt} &=& p_y^{can} {DV^1 \over Dt} = p^{can}_y (1 -{1 \over3} Ct^2)
= p_y, \\
{D^2 y \over Dt^2} &=& p_y^{can} {D^2 V^i \over Dt^2} = -Ctp_y^{can} 
= -Ctp_y = {[}p_y, p_0{]}.
\end{eqnarray}
We see that the geodesic momentum  $p_y$ describes the dynamics of
the gravitational field and its equation of motion -- the geodesic 
deviation equation -- is encoded in the commutator (\ref{kommp}).

%%%%%%%%%%%%%%%%%%%%%%%%%%%%%%%%%%%%%%%%%%%%%%%%%%%%%%%%%%%%%%%%%%%%%%%
{\bf 8. Discussion.}
The simple model of  a test particle moving in a two-dimensional Newtonian
spacetime allows us to clarify the physical meaning of
modified commutator algebra  (\ref{komm1})--(\ref{komm3})  in
the lowest approximation in  the Riemann coordinates 
$\{t,y\}$ and in the gravitational potential $\Phi$.   

All our model calculations can easily be generalized
to the case of a four-dimensional nonrelativistic Newtonian spacetime.
However, difficulties can emerge if we consider a relativistic
spacetime where the time coordinate is not absolute  and where a 
nonsingular metric tensor $g_{\mu \nu} $ exists.   

In the standard quantum mechanics, commutation relations are entirely
kinematical and dynamics is given by a Hamiltonian.
As distinct from this, since  the dynamics of the gravitational
field has been encoded in the coordinate transformation  (\ref{GR}), 
modified commutation relations  contain 
information about  kinematics as well as dynamics.

\bigskip
%%%%%%%%%%%%%%%%%%%%%%%%%%%%%%%%%%%%%%%%%%%%%%%%%%%%%%%%%%%%%%%%%%
{\bf Acknowledgements}
\medskip

The author  is grateful to  Eugen Paal for  enlightening 
discussions on nonassociative algebraic systems and critical reading
of the manuscript.    
This work was supported by the Estonian Science Foundation under
Grants No. 3308 and 3870.
%%%%%%%%%%%%%%%%%%%%%%%%%%%%%%%%%%%%%%%%%%%%%%%%%%%%%%%%%%%%%%%%%%%%%%%%%%%

\end{document}